# Online simulation powered learning modules for materials science

Samuel Temple Reeve[1], David M. Guzman[2], Lorena Alzate-Vargas[3], Benjamin Haley[3], Peilin Liao[3], and Alejandro Strachan[3]

1 Materials Science Division, Lawrence Livermore National Laboratory, Livermore, CA 94550

2 Condensed Matter Physics & Materials Science Division, Brookhaven National Laboratory, Upton, NY 11973

3 School of Materials Engineering and Birck Nanotechnology Center, Purdue University, West Lafayette, Indiana 47906 USA.

Simulation tools are playing an increasingly important role in materials science and engineering and beyond their well established importance in research and development, these tools have a significant pedagogical potential. We describe a set of online simulation tools and learning modules designed to help students explore important concepts in materials science where hands-on activities with high-fidelity simulations can provide insight not easily acquired otherwise. The online tools, which involve density functional theory and molecular dynamics simulations, have been designed with non-expert end-users in mind and only a few clicks are required to perform most simulations, yet they are powered by research-grade codes and expert users can access advanced options. All tools and modules are available for online simulation in nanoHUB.org and access is open and free of charge. Importantly, instructors and students do not need to download or install any software. The learning modules cover a range of topics from electronic structure of crystals and doping, plastic deformation in metals, and physical properties of polymers. These modules have been used in several core undergraduate courses at Purdue's School of Materials Engineering, they are self contained, and are easy to incorporate into existing classes.

## Introduction

Simulation tools spanning scales from electrons to structures are becoming an equal partner to experiments and theory in the field of materials science and engineering. Integrated computational materials engineering (ICME) has emerged as an important sub-field [1] and the US Materials Genome Initiative [2] calls for the use of simulation and data tools together with experiments to reduce the cost and time associated with the introduction of new materials. While the importance of simulations in materials research and development is beyond dispute, their power as a pedagogical tool has been not been exploited to the same degree. Two national surveys, published one year and ten years ago, detail the awareness, attitudes, and implementations of computational materials science and engineering (CMSE) in the classroom [3], [4]. While familiarity and degree to which inclusion of simulation tools is seen as important and has risen over the last decade, fully integrated use of simulation to enhance learning in core MSE courses is rare. An important exception is the University of Illinois at Urbana-Champaign efforts to incorporate simulations across their curricula [5]. While instructors are adopting online resources such as CES EduPack from Granta Design (https://www.grantadesign.com) and publishers have developed online resources including animations and illustrations (https://www.wileyplus.com/materials-science-and-engineering-an-introduction/), the use of high-fidelity simulations is not widespread outside of upper-level elective courses. One barrier towards the pervasive use of advanced simulations in the classroom is the lack of tools that are both easy to use and available at no or low cost. Powerful, open source tools for atomistic simulations, such as LAMMPS [6] for molecular dynamics (MD) and Quantum ESPRESSO [7], [8] for density functional theory (DFT), require a high level of expertise to install and operate, making them inaccessible to most instructors. Further, designing meaningful simulations and learning modules with these tools requires significant time and effort. Crucially however, incorporation of these tools in the undergraduate curriculum can help students develop a more intuitive understanding of the atomic and electronic level processes that govern materials behavior. For example, a molecular dynamics simulation of dislocation glide under various loading has been shown to be more powerful than static schematics found in textbooks, documented in several recent studies [9], [10].

To address this gap, we developed a series of online tools deployed in nanoHUB (https://nanoHUB.org) for electronic structure and atomistic simulations, designed with non-experts in mind (including instructors, students, and researchers). We exemplify their use via a series of learning modules to help students explore concepts related to electronic structure, mechanical properties of metals, and physical properties of polymers.

# Materials simulation tools in nanoHUB

nanoHUB is an open cyberinfrastructure for online simulation and collaboration [11] in education and research supported by the US National Science Foundation. The site is open and free of charge for both contributions and use, where users can perform fully interactive simulations from the web-browser. Simulation tool developers and authors of educational or research materials can use a self-serve process to publish their products (https://nanohub.org/resources/new) and running tools only requires registration or institutional credentials to log in to nanoHUB. The aforementioned survey of awareness and interest in CMSE resources ranked nanoHUB at the top in the context of educational use [4].

## DFT Materials Property Simulator

The DFT Materials Property Simulator tool (DFTmatprop) [12] was designed to introduce non-experts to electronic structure calculations with DFT by streamlining common calculations for a variety of materials. DFT is currently the method of choice for the vast majority of first principles electronic structure calculations in materials science due to its ideal combination of accuracy and computational efficiency. Most DFT codes solve the Kohn-Sham equations [13] to obtain the ground state electronic structure of atoms, molecules, or materials based on the Hohenberg-Kohn theorem [14]. From the ground state electronic structure of a material one obtains energy and derived quantities, such as forces and stresses, and also an approximate description of the single particle electronic structure (either energy levels in non-periodic systems or band structures in crystals). DFT is currently an indispensable tool in MSE, used in many applications.

## Under the hood.

The DFTmatprop tool combines an open source DFT code, Quantum ESPRESSO [7], [8], and a series of pre- and post-processing scripts used to prepare the simulations and extract the desired materials properties. Quantum ESPRESSO is a suite of independent and compatible Fortran codes for electronic structure calculations and materials simulations based on density functional theory that uses plane waves for the expansion of the electron wave functions and pseudopotentials to account for the electron-ion interactions. To simplify the user/tool interactions, we limited the pseudopotential library to high-quality norm conserving potentials generated with the Martins-Troullier method [15] spanning 85% of the periodic table. The electron exchange-correlation potential is calculated self-consistently using the local density approximation (LDA) or the generalized gradient approximation (GGA). The LDA and GGA implementations are limited to the Perdew-Zunger (PZ) [16] and Perdew-Burke-Ernzerhof (PBE) [17] functionals, respectively.

The DFTmatprop uses simple drop-down menus and free format dialog boxes to request the necessary information from the users to setup the input files and work directories to run Quantum ESPRESSO. The tool is capable of generating inputs to perform a wide range of simulations, from simple total energy calculations and full geometry relaxations to more sophisticated optical properties and transport simulations. DFTmatprop includes a set of post-processing Python scripts that generate total electronic density of states (DOS) plots, electron band dispersion diagrams (E-K diagrams), real and imaginary parts of the dielectric function as well as the refractive index and extinction and absorption coefficients discriminated along the x-, y- and z- vectors. DFTmatprop has been used by over 1300 unique nanoHUB users, who have collectively run nearly 20,000 simulations (https://nanohub.org/resources/dftmatprop/usage), and used in research publications (https://nanohub.org/resources/dftmatprop/citations).

## Pre-built calculations and advanced options.

When the tool is launched, users can select one of four pre-loaded property calculations, as shown in Figure 1(a), running purposeful DFT with even one click:

i) Equation of State: for a given system the tool applies a series of hydrostatic deformations to the unit cell and computes the total energy of the system at the relaxed ionic positions. From this set of data points, the post-processing scripts generate the energy vs. unit cell volume plot and computes the equilibrium volume and bulk modulus from a quadratic fit and the Murnaghan equation of state [18]. In the advanced options, the user can select the range of deformation percentage and the number of deformations.

ii) E-K diagram: the tool computes and generates plots of the total electronic DOS and the electron band dispersion. In the advanced options, the user can modify the energy range of the

DOS and E-K plots, as well as select the path in reciprocal space for the band dispersion plot. An example band structure output is shown in Fig. 1(b).

iii) Dielectric Constant and Optical Properties: the dielectric function (real and imaginary parts) is computed for semiconductors and insulators. Certain optical properties such as the refractive index and extinction and absorption coefficients are plotted along different directions. With advanced options, the user can specify the energy range, smearing and discretization of the dielectric function and optical properties plots.

iv) Thermoelectric Calculations: the tool computes the Kohn-Sham eigenvalues in dense and commensurate mesh in reciprocal space and outputs this data in a text file formatted to be used with the LanTrap tool [19] to compute thermoelectric transport properties using the Landauer formalism.

Additionally, DFTmatprop includes a set of pre-built structures spanning metals, semiconductors, and insulators. The list also includes some metallic surfaces along different crystallographic orientations and some semiconductor 2D materials such as transition metal chalcogenides.

For more advanced users, the tool has the flexibility to handle user-defined structures, selection of exchange-correlation functional, number of K-points in reciprocal space for the self-consistent and non-self-consistent field calculations, wave function and charge density kinetic energy cutoff, occupation and mixing schemes, and selection of ionic only or full geometry relaxation.

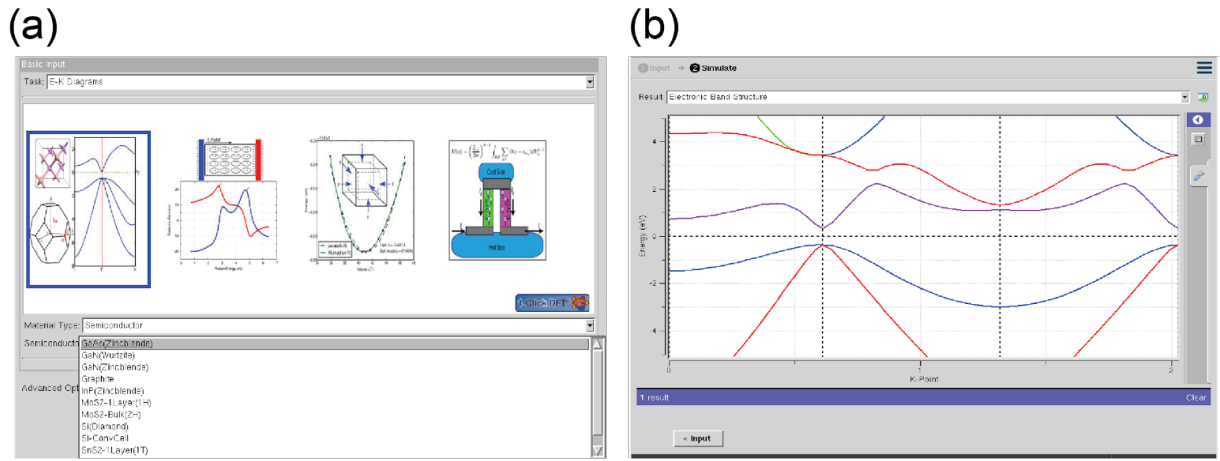

*Figure 1: (a) DFTmatprop front page showing main pre-built options, including band structures, dielectric properties, equations of state, and thermoelectric properties, as well as some of the available pre-built semiconductor structures. (b) Band structure of silicon plotted along high-symmetry points in the Brillouin zone.*

# Nanomaterial Mechanics Explorer

The Nanomaterial Mechanics Explorer (NanoMatMech) [20] was designed in the same spirit at the DFTmatprop, to introduce those new to MD simulations to predictions of thermo-mechanical properties of various materials. Following the pioneering work of Alder, Wainwright [21] and others over 60 years ago, MD simulates the dynamics of atoms using classical equations of motion. Central to MD simulations is the method used to compute atomic interactions. While these can be computed from electronic structure calculations (often DFT), such calculations are computationally intensive and often force fields (also called interatomic potentials) are used instead. Force fields provide an expression for the total energy of an atomic system in terms of the atomic positions from which forces can be calculated. The NanoMatMech tool enables users to select from a large variety of force fields by connecting to the openKIM [22] repository. The main output of an MD simulation is an atomistic trajectory from which thermodynamic quantities and structural information can be extracted. Thus, MD enables the investigation of complex processes on systems containing up to billions of atoms, although most simulations are significantly smaller. As with DFT, MD has been used to study a wide array of materials and properties.

## Under the hood.

The NanoMatMech tool combines the open source MD code LAMMPS [6] with custom scripts to initialize the simulations and extract the results and VTK for visualization. LAMMPS enables MD simulation on both personal computers and massively-parallel supercomputers. In order to run LAMMPS, an atomic structure, force fields, and simulation test are all required. For NanoMatMech, the instructions to build structures and tests primarily use LAMMPS, as well as the Atomsk toolkit [23], with analysis similarly done through both LAMMPS and the OVITO Python interface [24]. Force fields are included within the tool and through on-the-fly connection to the OpenKIM database [22]. Only embedded atom method (EAM) force fields [25] are included in the tool due to their balance of accuracy and computational efficiency for metals; the EAM includes short-range repulsion, pair attraction, and forces representing the insertion of an atom into the electron cloud of surrounding atoms. Through OpenKIM, over 25 elements and 20 alloys are available to simulate with over 80 force fields. Python scripts are used to extract the LAMMPS results to convert from standard thermodynamic properties to the mechanical outputs of interest and show atomic structures. NanoMatMech has been used by over 1100 nanoHUB users, has launched over 13,000 simulations (https://nanohub.org/resources/nanomatmech/usage); citations in scientific and educational research articles can be found at https://nanohub.org/resources/nanomatmech/citations.

## Pre-built calculations and advanced options.

Four main sections are built within the quick-start section of the tool, shown in Figure 2(a): crack propagation, dislocation dynamics, phase transitions, and tensile testing. Each of the four offers pre-built tests varying one or two variables of physical interest; for example, changing the crystal structure (FCC and BCC) and temperature enables exploring ductile vs. brittle failure for crack propagation. Multiple crystal structures and dislocation types (edge and screw) are available for dislocation dynamics, transition type (melting and martensitic) and boundary conditions for phase transitions, and multiple materials and crystal orientations for nanowire tensile tests.

Therefore, with 1-3 clicks for each simulation, meaningful comparisons of materials concepts are made possible. In addition, many other details are exposed through the advanced options: material, including force field, lattice, and composition; structure, including total size, defects, and orientation; physical test, including temperature, strain, and run time; and selection of outputs. Example outputs including atomistic visualization and curve outputs for each of the four sections is shown in Fig. 2(b).

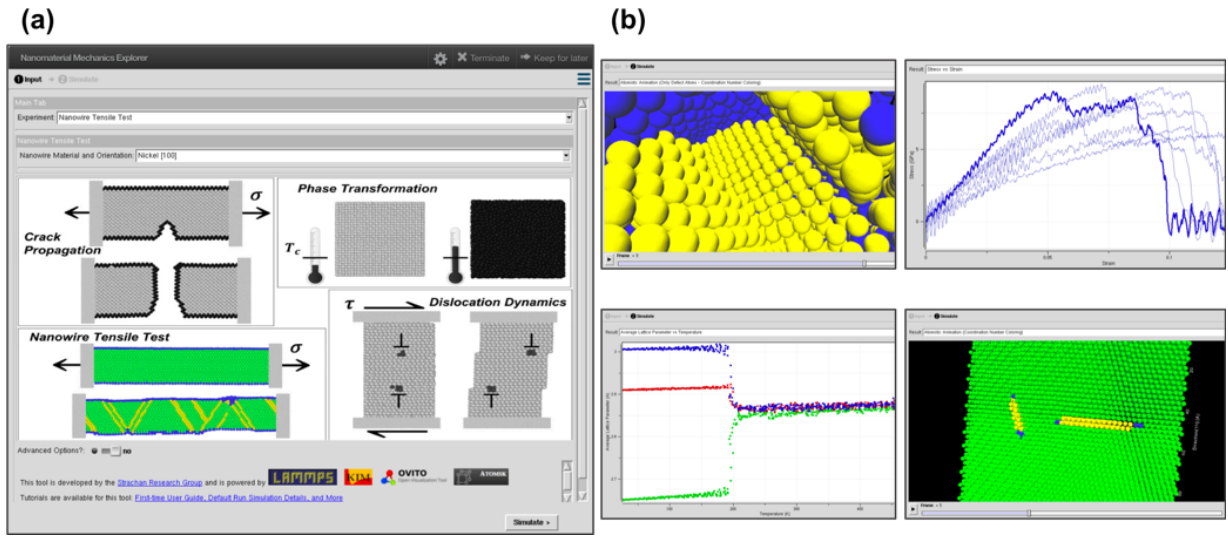

*Figure 2: a) NanoMatMech front page showing main pre-built options, including crack propagation, tensile tests, dislocation dynamics, and phase transformations. b) Examples of NanoMatMech results: zoomed view of dislocations within a nanowire tensile test, comparison of multiple crack propagation runs of varying temperatures, martensitic transformation during cooling showing changes in lattice parameter, and screw dislocations interacting.*

## Polymer Modeler

MD simulations of polymers can also provide valuable information for science and engineering. A key challenge in molecular simulations of polymeric systems is to start with an accurate representation of the polymer's molecular structure. This is particularly true for polymers with large molecular weights as their relaxation timescales are beyond what is possible to achieve with MD. The Polymer Modeler [26] tool is designed to create such structures for linear polymers using an in-house Monte Carlo (MC) algorithm [27]. PolymerModeler can follow the amorphous building step with MD simulations at a chosen temperature to further relax the structure. Following this thermalization, MD simulations can be performed to characterize thermo-physical and mechanical properties.

### Under the hood.

PolymerModeler is based on a custom continuous configurational biased Monte Carlo (CCBMC) [28] code to build atomic structures, can type force fields, and can connect to LAMMPS for MD simulations. Users can run the structure builder, select a force field, and and perform a MD simulations with LAMMPS without knowledge of the details of each piece of the simulation, such as input file formats. Monomers can be specified in PDB, XYZ, or z-matrix formats. In addition to CCBMC, other chain builder algorithms are available for understanding and teaching, e.g. freely rotating chains. Both DREIDING [29] and ReaxFF [30] force fields are available, for fixed bond and reactive simulations, respectively; partial atomic charges are determined by the Gasteiger approach and both particle-particle particle-mesh and Ewald methods are accessible for long-range electrostatic interactions. PolymerModeler has been used by over 1700 visitors to nanoHUB, together amassing more than 44,000 simulation runs (https://nanohub.org/resources/polymod/usage), and has been cited within numerous research publications (https://nanohub.org/resources/polymod/citations).

### Pre-built calculations and advanced options.

PolymerModeler runs with sensible default values so that novice users can run a reasonable simulation in a few clicks, but the details are readily accessible to expert users who want to control the details of the MC build or the many MD options, such as time step or cooling rate. Users start by selecting one or more monomers, either from a preexisting list or uploading their own, as shown in Figure 3(a). Then, they select the number of chains, number of monomers per chain, and whether to build homo- or co-polymers. Following these molecular-level parameters, the users can specify density of the simulation cell and several other parameters that control the build of the system. Finally, users can specify whether to only build the polymer structure or follow the build with an MD simulation using LAMMPS. Several options enable users to design simulations to thermalize the system and perform calculations like cooling to compute the glass transition of the polymer. Main outputs include the LAMMPS data file to continue running simulations outside the tool, Fig. 3(b), and atomistic visualization of the created structure, Fig. 3(c).

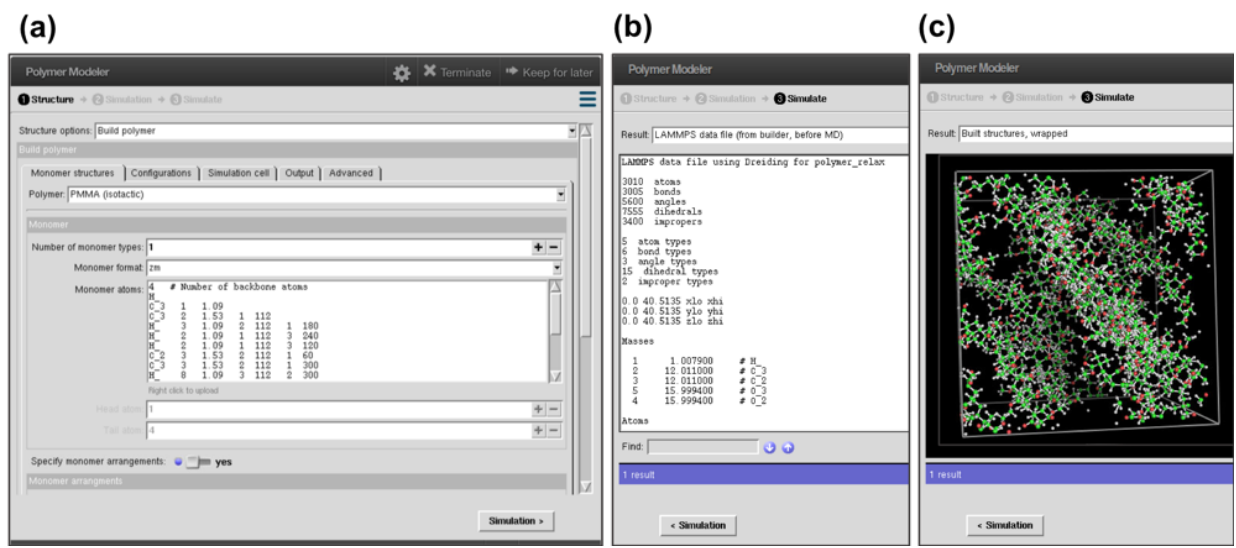

*Figure 3: a) PolymerModeler front page showing main options for polymer building and details for monomer structure, in addition to the polymer relaxation tab. b) PolymerModeler produced LAMMPS data file from structures generated using a CCMBC algorithm and atomic types*

*determined according to the chosen force field. c) Interactive atomic structure of an isotactic PMMA system generated by PolymerModeler*

# Learning modules

This section describes several simulation-powered learning modules designed for and implemented in various core MSE courses at Purdue. They include electronic structure and MD calculations to explore electronic, mechanical, and physical properties and processes in semiconductors, metals, and polymers. These modules have evolved through many instructors and courses, as have the simulation tools they rely on.

## Electronic structure and binding in Si

Quantum mechanics is a challenging subject to introduce to undergraduate engineering students, yet it is critical to understand bonding and structure of materials, electronic properties and a long list of other processes. One particularly complex concept for students is the electronic band structure of crystals. Band structures of real materials (spaghetti diagrams) can be hard to understand, but are important to for electronic and optical properties and the operation of technologically important devices.

### The learning objectives

of this module are to understand how the band structures form starting from an isolated atom and the meaning of dispersion (dependence on k). In addition, students explore how the relative importance bonding-antibonding splitting and s-p splitting controls the metallic or semiconducting character of group IV elements.

### Module overview.

In the module "*Binding and electronic structure of silicon*" (https://nanohub.org/resources/30102) students explore increasingly complex electronic structures using DFT in the DFTmatprop tool [12]. An additional module which uses a previous tool is also available (https://nanohub.org/wiki/LearningModuleSiliconBandstructureDFT). The first step is to calculate the electronic structure of an isolated Si atom (actually a collection of non-interacting Si atoms). This is obtained by performing a DFT calculation on a diamond crystal structure (using a two atom unit cell with an FCC Bravais lattice) but with an extremely large lattice parameter, 20A. Under these conditions, atomic interactions are negligible and the electronic structure is characterized by two energy levels, the 3s and 3p states of Si. This is relatively easy for the students to understand. The second step, see Figure 4(b), is to leave the lattice parameter unchanged but bring the two atoms together to create a Si dimer. Under these conditions, the s and p energy levels split into bonding and antibonding states. There is no dispersion (dependence on wave vector k) as there is no interactions between molecules in subsequent cells. Finally, we go back to the diamond crystal with a 20A lattice parameter and slowly decrease the lattice parameter down to the equilibrium value. The students analyze: i) how the s and p states transform into bands (with dispersion this time), ii) how at a given lattice parameter the bonding anti-bonding splitting matching the s-p splitting and the s and p bands merge and iii) how a gap appears within the band resulting in a semiconductor, see Figure 4.

## Audience and classroom use.

This module was developed in 2011 and used since then as a homework assignment in elective courses on CMSE at Purdue. Since 2015 the module has been used as an in-class demonstration for Purdue's MSE 270 "Atomistic Materials Science", a core sophomore-level class with enrollment between 50 and 80 students.

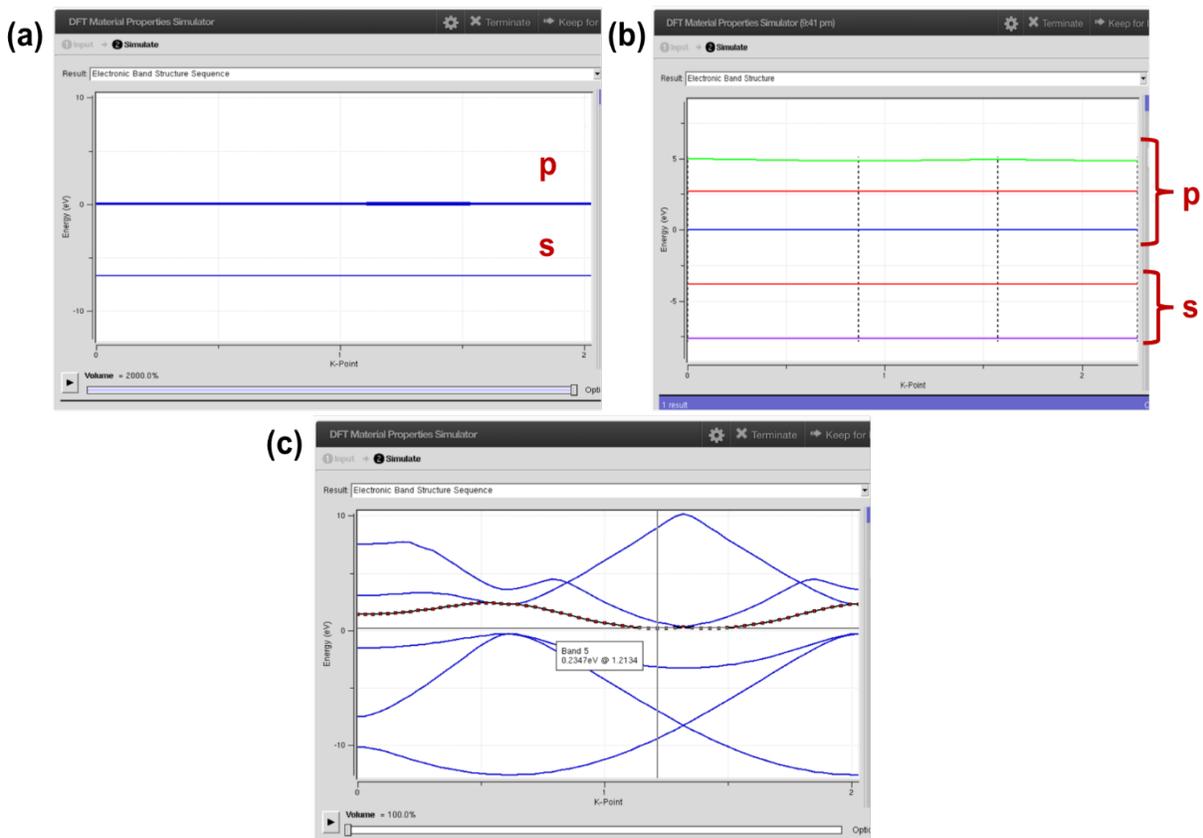

*Figure 4: Band structures in silicon for a) isolated atoms, b) a dimer, and c) diamond cubic crystal. Here, a) and c) are from the initial and final volumes of the system, with many points in between showing the process of energy levels developing into bands and the formation of a band gap. Bonding and anti-bonding splitting is shown in b).*

## Doping in semiconductors

Silicon is a widely used semiconductor in optical and electronic devices. Its carrier concentration and electronic structure are often tailored by doping for the specific applications. Some dopants (typically group V elements) can donate electrons to the crystal's conduction bands, transforming it into an n-type semiconductor; on the other hand, other dopants (typically group III) can accept electrons resulting in holes in Si's valence bands, making it a p-type semiconductor. We designed a learning module for students to explore Si doping using DFT calculations, students can explore, hands-on, how different elements affect the electronic band structure.

### The learning objectives

of this module are: i) explore band structures and electronic DOS, ii) identify occupied/unoccupied states in band structures and DOS plots and differentiate various types of solids (metals, semiconductors or insulators, as well as direct vs indirect band gap materials, iii) compare DOS of pure and doped Si and explain the difference.

### Module overview

In the module "*Band Structure for Pure and Doped Silicon*" (https://nanohub.org/resources/29319) students perform online DFT simulations using DFTmatprop [12] to compute band structures and DOS for pure crystalline Si and Si doped with various elements. Once the simulations finish, students analyze the DOS and band structure results to determine the electronic nature of the system (metallic, semiconducting or insulating). In addition, they will study variation in DOS for doped semiconductors, to gain understanding in the mechanism of different types of doping.

The prelab lecture consists of a brief introduction to DFT, followed by information on band structures and DOS plots. We discussed the concepts of k-points, how the bands run, band dispersion, and occupied/unoccupied states in band structures and DOS. We introduce the tool and discuss the physical meaning of important input parameters and how to modify input structures.

Students run three simulations: pure Si, Ga-doped Si, and P-doped Si. Each simulation takes about 10 minutes. First, students investigate pure crystalline Si. The students were asked to visualize Si crystal structure, label the conduction band minimum (CBM) and valence band maximum (VBM) of Si band structure, and identify whether Si is direct or indirect band gap material. Next, students were asked conceptual questions about electron counts in Ga, Si, and P. Then they are requested to modify input structures to introduce Ga or P dopant into the Si crystal. After obtaining the simulation results, students will compare the Fermi level, CBM, VBM, and overall shape of DOS among pure Si, Ga-doped Si, and P-doped Si. They must apply their lecture/textbook knowledge of p-type/n-type doping to explain changes in DOS plots. Figure 5 shows the concept for this module, with the relevant tool input for P-doped Si, as well as the output DOS for both pure and doped Si.

### Audience and use in the classroom.

This module was developed for and used in Purdue's MSE 370 "Electrical, Optical, and Magnetic Properties of Materials" core junior-level course. The module has been used two semesters to date. The full implementation includes a 30 min pre-lab lecture and a lab report.

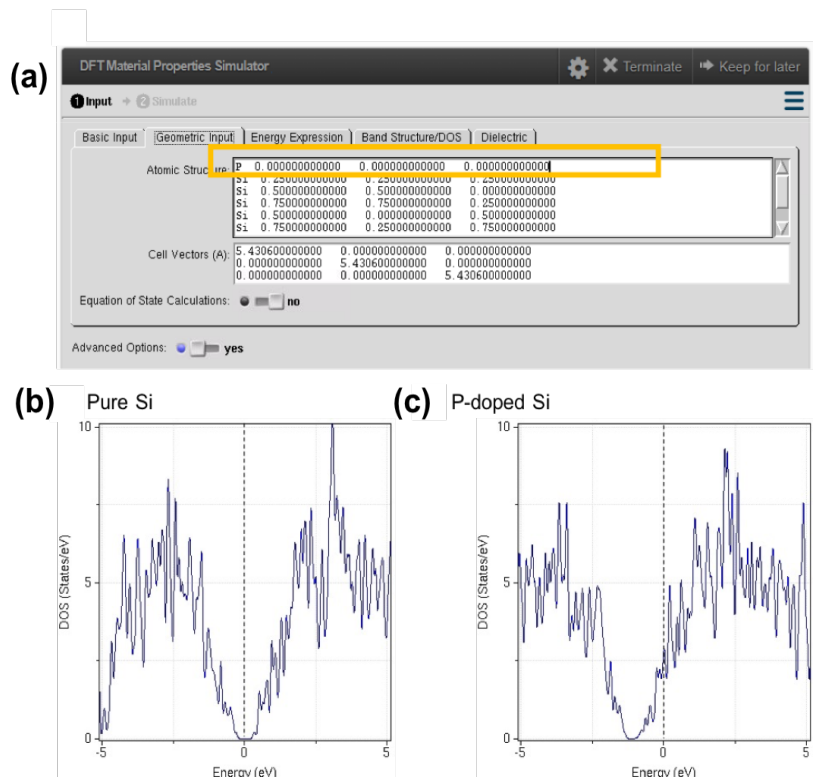

*Figure 5:* (a) Screenshot of the DFT Materials Properties Simulator tool illustrating changes for Atomic Structure necessary for the dopant study. (b) DOS image output for pure Si and (c) DOS image output for phosphorus doped Si. The Fermi level is set to be at 0 eV in the DOS plot (highlighted by black vertical dashed line).

## Mechanisms of plastic deformation in metals

Under most conditions, plasticity in metals is governed by the creation, motion, and interaction between dislocations with each other and other defects. We hypothesized that MD simulations of plastic deformation combined with powerful visualization of the resulting trajectories could help students develop a stronger mechanistic understanding of plastic dislocations, dislocation motion, slip planes and the concept of critical resolved shear stress. A study of student performance with an early version of this module confirmed our hypothesis [9]. Further, we speculated that such a learning module would enable students to compare the difference in response between a nanoscale wire and macroscopic polycrystalline samples, bridging topics they learn in lecture classes and lab experiments. Student performance and understanding of these concepts was explored as the module evolved in Ref. [10] and references therein.

### The learning objectives

of this module are: i) compare and contrast the characteristic features of stress-strain curves of a defect-free nanoscale sample and a macroscopic polycrystalline one, ii) connect the atomic structure and processes to features in the stress-strain curve iii) explain the lack of hardening in the nanowire, iv) identify slip planes and explain plasticity from the atomic scale.

## Module overview

In the module "*Tensile Testing Laboratory: Nanoscale and Macroscale Metal Samples*" (https://nanohub.org/resources/23130) students perform MD simulations of the deformation of a metallic nanowire using NanoMatMech [20], together with an experimental tensile testing component. Once the simulations finish, students analyze the resulting stress-strain curves and calculate quantities such as Young's modulus, yield strength, and hardening. In addition, they analyze atomistic snapshots of the system during various stages of deformation to correlate atomic-level processes to features in the stress-strain curves. For the lab course, a guide was provided highlighting the above learning objectives, detailed instructions, and lab report requirements. Based on previous iterations of the lab many specific points were highlighted. An additional module "*Nanoscale tensile testing with MD*" (https://nanohub.org/resources/30014) contains only the MD portion of the joint experimental/computational module. It was also modified with new questions and additional simulations requiring student modification of advanced inputs for greater understanding of the physical inputs. Figure 6 shows the stress-strain response of the tool together with inset atomistic initial and final structures for one case.

## Audience and use in the classroom

This module was developed and implemented for a sophomore-level course, Purdue's MSE 235 "Materials Properties Laboratory", connected with an existing experimental tensile testing lab. The original version of the laboratory module was deployed in 2010 (https://nanohub.org/topics/LearningModulePlasticityMD) and the current module has been used by MSE 235 in small group lab projects since 2015. Enrollment in Purdue's MSE 235 has recently been between 50 and 80 students.

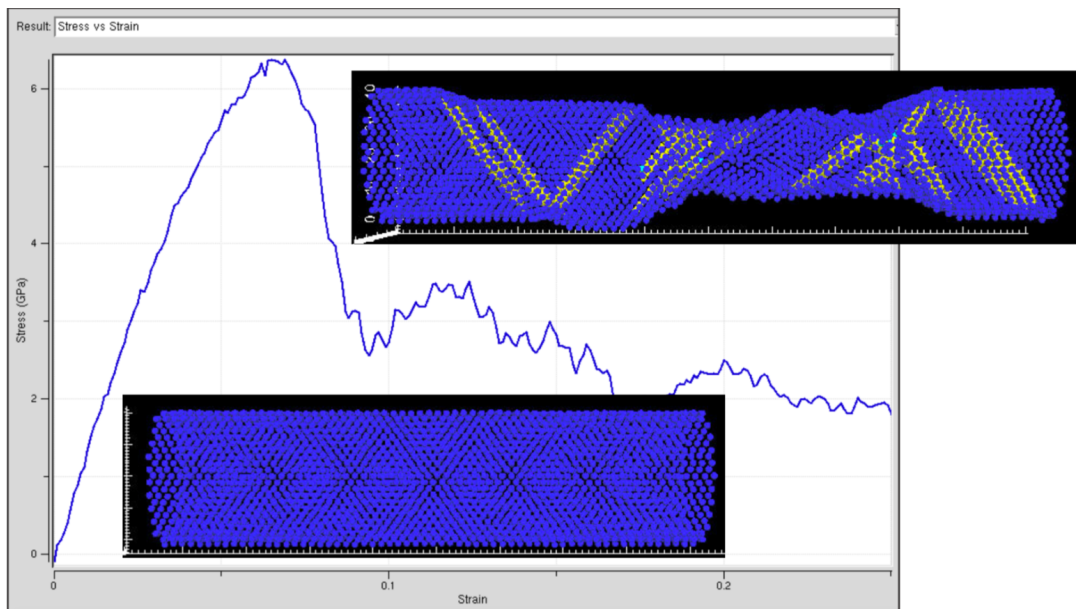

*Figure 6: Screenshot of the Nanomaterial Mechanics Explorer tool results for a [110] orientation nickel nanowire. The stress-strain response output is shown with inset atomic structure for the*

*initial and final (25% strain), where the latter highlights dislocations in yellow with significant plasticity and necking.*

# Heat of fusion of polyethylene

Semicrystalline polymers consist of crystalline and amorphous phases and are widely used in everyday (e.g. plastic bottles and containers) and advanced aerospace applications. The degree of crystallinity of these materials affects their thermo-mechanical properties and such structure-properties relationships are covered in several introductory and advanced MSE courses. The degree of crystalline of polymer samples can be obtained experimentally from differential scanning calorimetry (DSC) experiments where the total energy required to melt the material can be obtained. In addition to the experimental observable, calculating the degree of crystallinity requires knowledge of the heat of fusion of the system. This quantity is not easily extracted experimentally; however, MD simulations can provide a reliable value.

## The learning objectives

of this module are for students to: i) visualize and explore the molecular structure of crystalline and glassy polymers, and 2) understand the concept of heat of fusion and calculate it using MD simulations.

## Module overview

In the module "*Calculating the heat of fusion of polyethylene using Polymer Modeler*" (https://nanohub.org/resources/29339) students calculate the heat of fusion of a polyethylene sample from MD of the crystalline and amorphous samples. Students perform the MD simulations with PolymerModeler [26] using the pre-built and thermalized samples. The sample sizes have been chosen so each MD simulation takes about 45 minutes as a compromise between expedience and accuracy. In addition to calculating the heat of fusion used to compute the percent crystallinity in the experimental lab, students visualize both crystal and amorphous samples during the simulation to develop a better understanding of the molecular structures of polymers. The tutorial includes instructions on how to select the structures, how to setup and run the simulation and how to extract the outputs needed and important information for calculating the heat of fusion. Figure 7 shows the visualized atomic structure comparing crystal and amorphous samples, as well as their output energies.

## Audience and use in the classroom

This module was also developed for and used within the sophomore-level course at Purdue, MSE 235 "Materials Properties Laboratory", in Spring 2019. The module was designed to enhance an existing lab where students use differential scanning calorimetry with semi-crystalline polymers.

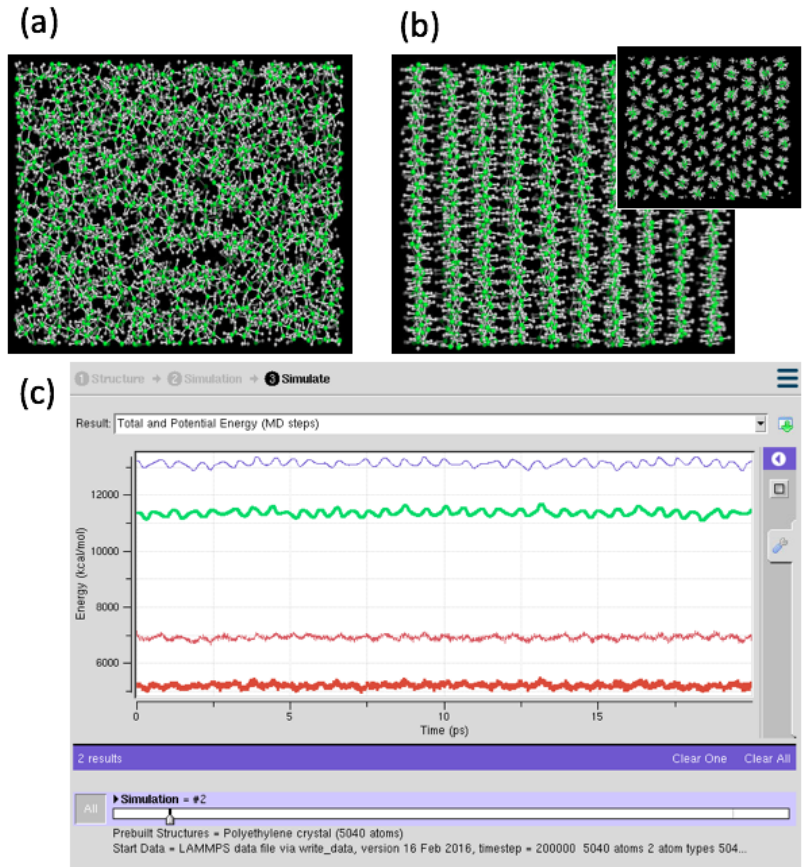

*Figure 7: (a) Snapshot of the pre-built amorphous configuration as seen in PolymerModeler. (b) Snapshot of crystalline configuration from PolymerModeler and (c) Screenshot of total and potential energy vs time of comparing crystalline and amorphous polyethylene samples in PolymerModeler results after MD simulation.*

## Additional modules

The modules "*Electronic structure and spin of the O atom*" (https://nanohub.org/resources/21956) and "*Electronic structure and spin in O2 molecule*" (https://nanohub.org/resources/21954) use DFT to explore the difference in energy and structure between O and O2 in their singlet (spin 0) and triplet (spin 1) states. The goal of the module is for students to explore exchange interactions and a quantitative understanding of its importance in determining the ground state of atoms and molecules.

The module "*Melting with MD*" (https://nanohub.org/resources/30012, https://nanohub.org/resources/22025) provides step by step instructions to setup and perform MD simulations to study melting in metals. Students construct a supercell and heat the sample until melting is observed. Atomistic structures of the crystalline and liquid samples can be explored via 3D visualization. In addition, students build a nanoparticle and explore melting temperature depression in nanoscale materials.

The module "*Dislocation structure and propagation with MD*" (https://nanohub.org/resources/30006) uses MD simulations to show students what dislocations look like atomistically and, most importantly, how they move. Both FCC and BCC crystals, with both pure edge and screw dislocations are accessible with the correct strain applied for glide. When students change that applied strain, new dislocations nucleate and stress magnitudes change dramatically.

The module "*Ductile and brittle crack propagation in metals with MD*" (https://nanohub.org/resources/30008) uses MD simulations to study crack propagation in FCC and BCC metals. By changing temperature students can visualize the atomic difference between brittle crack opening and a ductile case where dislocations are emitted from the crack tip resulting in its blunting and arresting catastrophic failure.

The module "*Martensitic transformations with MD*" (https://nanohub.org/resources/30010) uses MD simulations to show the atomic process of transformation in common shape memory alloy NiTi and other metals. The change in crystal structure and abrupt transition is clear and connections are drawn to the industrially important transition in steels.

All the tutorials and tools discussed here and others of interest to the MSE community can be found in the Materials Group in nanoHUB: https://nanohub.org/groups/materials.

## Discussion and conclusions

Modeling and simulations can play an important role in the education of MSE students. While traditionally simulations were introduced in technical elective or graduate courses, the increasing reliance on simulations in the field calls for the introduction of these tools in core undergraduate courses. In addition, powerful simulations have a significant pedagogical potential and can provide students with hands-on experience generating and analyzing data and visualization can help students develop a more intuitive understanding of complex topics. In this paper we introduced a series of simulation tools designed for non-experts and available online at no-cost to the user with associated learning modules we have used in several MSE courses at Purdue. We believe the modules are easy to incorporate in existing classes and do not require the instructor to be an expert in simulations. Finally, we believe MSE students and instructors would benefit from additional modules and homework assignments being openly shared online.

## Acknowledgements

This work was partially supported by the US National Science Foundation EEC-1227110. This work was performed in part under the auspices of the U.S. Department of Energy by Lawrence Livermore National Laboratory under Contract DE-AC52-07NA27344. Computational resources and staff support from nanoHUB.org is gratefully acknowledged.